# Statistical evaluation of a long-memory process using the generalized entropic Value-at-Risk

Hidekazu Yoshioka[a] *, Yumi Yoshioka[a]

[a] Shimane University, Nishikawatsu-cho 1060, Matsue 690-8504, Japan
* Corresponding author: E-mail: yoshih@life.shimane-u.ac.jp, TEL: +81 852 32 6541

**Abstract** The modeling and identification of time series data with a long memory are important in various fields. The streamflow discharge is one such example that can be reasonably described as an aggregated stochastic process of randomized affine processes where the probability measure, we call it reversion measure, for the randomization is not directly observable. Accurate identification of the reversion measure is critical because of its omnipresence in the aggregated stochastic process. However, the modeling accuracy is commonly limited by the available real-world data. One approach to this issue is to evaluate the upper and lower bounds of a statistic of interest subject to ambiguity of the reversion measure. Here, we use the Tsallis Value-at-Risk (TsVaR) as a convex risk measure to generalize the widely used entropic Value-at-Risk (EVaR) as a sharp statistical indicator. We demonstrate that the EVaR cannot be used for evaluating key statistics, such as mean and variance, of the streamflow discharge due to the blowup of some exponential integrand. In contrast, the TsVaR avoids this issue because it requires only the existence of some polynomial, not exponential moment. As a demonstration, we apply the semi-implicit gradient descent method to calculate the TsVaR and corresponding Radon–Nikodym derivative for time series data of actual streamflow discharges in mountainous river environments.

**Statements and declarations**

*Acknowledgments*

This work was supported by Japan Society for the Promotion of Science (22K14441, 22H02456) and Environmental Research Projects from the Sumitomo Foundation (203160).

*Availability of data*

Data will be available upon reasonable request to the corresponding author.

*Competing interests*

The authors have no competing interests.

*Author contributions*

Hidekazu Yoshioka: Conceptualization, Methodology, Software, Formal analysis, Data Curation, Visualization, Writing–original draft preparation, Writing–review and editing, Supervision, Project administration, Funding acquisition

Yumi Yoshioka: Data Curation, Visualization, Writing–review and editing

1. **Introduction**

**1.1 Research background**

Environmental management and restoration require a deep understanding of physical, biological, and chemical processes that vary dynamically and often stochastically (Phillips, 2022; Zeng et al., 2022). In particular, streamflow environments, which act as pathways and habitats for aquatic species, have been critically degraded by anthropogenic disturbances, including flow alteration for hydropower generation (Hough et al., 2022), biodiversity losses due to intermittent flow (Darty et al., 2022), and nonpoint source pollution from agriculture (Chen and Wang, 2022).

Stochastic process models are widely used for the modeling and risk analysis of dynamical systems such as streamflow environments. Many studies have considered streamflow discharge from the viewpoint of jump-driven stochastic processes at both the local (Wang and Guo, 2019; Yoshioka et al., 2023) and regional scales (Durighetto et al., 2022; Bertassello et al., 2022). Accelerated erosion due to changes in land use requires estimation of highly stochastic turbidity dynamics in a streamflow environment as a principal water quality index (Steffy and Shank, 2018). Probabilistic analysis can be used to derive a flow–rating curve from hydrological time series data (Hrafnkelsson et al., 2021). Because stochastic process models are conceptual rather than mechanistic, they inherently contain modeling errors due to simplification and/or coarse-graining. Further, unpredictable changes in climate can result in unintended model distortion after a tipping point is reached (Mastrantonio et al., 2022). This implies that a stochastic process model is always at risk of ambiguity due to structural errors in coefficients and parameters (Herrera et al., 2022; Gupta and Govindaraju, 2023). The modeling and analysis of stochastic processes subject to ambiguities have long been studied in economics and other related fields, where the ambiguity is probabilistically measured by using the Radon–Nikodym derivative between the benchmark and distorted models (Anderson et al., 2003; Dowd et al., 2008). In these fields, many studies have applied risk measures to real-world time series data (Imai, 2022; Tsang et al., 2022). The mathematical universality of model ambiguity is applicable to stochastic processes not only in economics but also in other research fields, including environmental management and restoration.

Another key issue for stochastic process models is subexponential or long memory (Beran et al., 2013), where the memory refers to the tail behavior of the autocorrelation function. A stochastic process is considered to have a long memory if its autocorrelation function is not integrable (e.g., the autocorrelation function has the polynomial decay $\tau^{-a}$ ($a \in (0,1]$) where $\tau$ is the time lag). The hourly data of the streamflow discharge in mountainous rivers have a long memory (Yoshioka, 2021; Yoshioka et al., 2023). The long memory of a streamflow discharge has been explained as the presence of watershed runoff processes with multiple time scales, which emerge as a superposition of exponential correlations with different relaxation rates (Mudelsee, 2007; Ranjram and Craig, 2022). Therefore, the stochastic process model of a streamflow environment should be considered as a long-memory process subject to ambiguity. However, this subject has not been studied in depth to the best of our knowledge.

**1.2 Objective and contribution**

In this paper, our objective is to introduce optimization problems that can consistently determine the lower and upper bounds of the mean, variance, and un-normalized skewness and kurtosis of streamflow discharge data. The un-normalized skewness and kurtosis refer to the third- and fourth-order centered moments without non-dimensionalization by the variance. The streamflow discharge is a fundamental indicator of a streamflow environment and an indispensable background driver of water quality indices (Diamond et al., 2022; Laroche et al., 2022) and habitat suitability indices for aquatic fauna (Qiu et al., 2022; Rinaldo and Rodriguez-Iturbe, 2022). Its statistical evaluation is therefore an indispensable step toward the water environmental assessment.

We represent the streamflow discharge as a stochastic process comprising the superposition of Ornstein–Uhlenbeck processes, which we refer to hereafter as the supOU process (Barndorff-Nielsen, 2001). The supOU process is an aggregation of infinitely many mutually-independent OU processes with different reversion speeds $r > 0$. The aggregation is mathematically justified in the sense of Lévy bases or independently scattered random measures (Pedersen, 2003; Barndorff-Nielsen and Stelzer, 2011) that serve as idealized models of infinite-dimensional noises. The probability measure for aggregating the reversion speed $\pi(\mathrm{d}r)$, which we call the reversion measure, is a key component of the supOU process because it appears in the statistical moments (mean, variance, skewness, kurtosis) and autocorrelation function. Specifically, the reversion measure mostly appears through the inverse moment of reversion speeds:

$$R = \int_0^{+\infty} \frac{1}{r} \pi(\mathrm{d}r). \tag{1}$$

This means that the reversion measure should be sufficiently regular near the origin $r = 0$. The Gamma-type measure is widely used as a convenient model (Fasen and Klüppelberg, 2007; Yoshioka, 2021) because it gives rise to closed-form statistical moments and an autocorrelation function with subexponential decay:

$$\pi(\mathrm{d}r) = \frac{1}{\Gamma(\alpha)\beta^\alpha} r^{\alpha-1} \exp\left(-\frac{r}{\beta}\right) \mathrm{d}r, \quad r > 0 \tag{2}$$

where $\alpha > 1$ and $\beta > 0$, and $\Gamma$ is the Gamma function. Similar aggregated models have been analyzed for continuous-time cases (Randrianambinina and Esunge, 2022) and for discrete-time cases where there may be a correlation among elements of aggregated processes (Beran et al., 2020; Vera-Valdés, 2020).

We can assume that the reversion measure $\pi$ identified by some method is ambiguous and that the true or most reasonable measure is the perturbation $\phi(r)\pi(\mathrm{d}r)$ with some positive function $\phi$ so that $\phi(r)\pi(\mathrm{d}r)$ becomes a new probability measure. The coefficient $\phi$ can be understood as a Radon–Nikodym derivative or its function between the benchmark and distorted models. The entropic Value-at-Risk (EVaR) (Ahmadi-Javid, 2012) is a widely used risk measure owing to its analytical tractability and close relationship with other existing risk measures. Applications of EVaR include economics (Chan et al., 2019), mechanical engineering (Wasserburger et al., 2020; Watanabe et al., 2022), and portfolio optimization (Yu and Sun, 2021).

The Tsallis formalism for risk assessment has become a popular approach owing to its greater flexibility than classical entropic approaches and its close relationship to physical phenomena and efficient machine learning schemes (Ma and Tian, 2021; Kumbhakar, and Tsai, 2022; Leleux et al., 2021; Cao et al., 2019). In this paper, we use the Tsallis Value-at-Risk (TsVaR) (Zou et al., 2022) as a recently introduced generalization of the EVaR. There are two reasons why we use the TsVaR rather than the EVaR. First, the TsVaR reduces to the EVaR with a suitable shape parameter, so a theory based on the former would be more versatile than one based on the latter. Second, the EVaR is not applicable to long-memory processes such as the supOU process because it requires the existence of the integral of $\exp(cr^{-1})\pi(\mathrm{d}r)$ with some $c > 0$, which turns out to diverge and hence is ill-defined. In contrast, TsVaR requires only the existence of a polynomial moment such as $r^{-1/(1-q)}\pi(\mathrm{d}r)$ with some $q \in (0,1)$, which is less restrictive. Such theoretical issues have rarely been discussed with real data. Indeed, inverse moments as in (1) seem to be less frequent in moments of monomials and polynomials.

TsVaR has been considered for the worst-case overestimation of a random variable of interest because the main motivation for its development was the evaluation of tail risks (Zou et al., 2022). However, we are considering not only the upper bound of a statistic but also the lower bound. We thus propose another TsVaR for the worst-case underestimation problem. Our proposed approach can be used to consistently evaluate both the lower and upper bounds of statistics for the supOU process. In addition, it does not require the use of Hamilton–Jacobi–Bellman type equations for evaluating worst-case statistics (Balter et al., 2021; Yoshioka and Yoshioka, 2022), where the proposed setting becomes very high- or infinite-dimensional partial differential equations that are difficult to handle. Because we do not use these equations, our approach completely avoids this issue.

For the numerical computation of the TsVaR, we exploit the fact that they are defined through a convex optimization problem that can be handled by a semi-implicit gradient descent method. Our computational approach is not completely new, but providing computational examples of TsVaRs is important because no or at most only a few such studies exist to the best of our knowledge. In addition, we propose a computational method for obtaining the Radon–Nikodym derivatives corresponding to the worst-case overestimation and underestimation of TsVaRs, which is based on exploiting the duality between TsVaRs and some nonlinear expectations. As a demonstration, we apply our proposed approach to identifying the supOU processes of streamflow environments in Japan and evaluating the TsVaRs.

Thus, our paper contributes to the theory, computation of a generalized EVaR, with a particular emphasis on its application to streamflow environments. The rest of the paper is organized as follows. **Section 2** reviews the fundamental properties of the supOU process and TsVaR. A TsVaR for the underestimation case is introduced, and the gradient method for computing TsVaRs and use of duality to obtain the Radon–Nikodym derivatives are presented. **Section 3** presents the application of our methodology to real-world data. **Section 4** concludes the paper and gives our perspective on the future prospects of our methodology. **Appendix A** in **Supplementary Material** contains proofs of propositions in the main text.

## 2. Mathematical model

### 2.1 SupOU process and its statistics

#### 2.1.1 Basic properties

The supOU process is an aggregation of mutually-independent OU processes having different reversion speeds with each other (Barndorff-Nielsen, 2001). The supOU process $X$ considered in this paper is represented by the following stationary stochastic integral at time $t \in \mathbb{R}$:

$$X_t = \int_0^{+\infty} \int_0^{+\infty} \int_{-\infty}^{t} \exp\left(-r(t-s)\right) \mu(\mathrm{d}s, \mathrm{d}z, \mathrm{d}r). \tag{3}$$

Here, $\mu$ is the Poisson random measure of the pure-jump type having the compensator $\mathrm{d}s \times v(\mathrm{d}z) \times \pi(\mathrm{d}r)$ with $v$ the Lévy measure of a subordinator governing the law of jump size $z$, which is of a bounded-variation independent increment jump process, such that (Section 11.2 of Hainaut (2022))

$$\int_0^{+\infty} \min\{1, z\} v(\mathrm{d}z) < +\infty \quad \text{but not necessarily} \quad \int_0^{+\infty} v(\mathrm{d}z) < +\infty. \tag{4}$$

The latter boundedness condition is satisfied if $v$ is of a compound Poisson type. The reversion measure $\pi$ is a probability measure and therefore satisfies the natural requirement $\int_0^{+\infty} \pi(\mathrm{d}r) = 1$. It is assumed to comply with the boundedness of the inverse moment (1). The integration of (3) is carried out for $s$, $z$, $r$ from inside to outside. The parameter $r > 0$ in the integral represents the reversion speed distributed according to the reversion measure $\pi$. The supOU process is viewed as a large-population limit of mutually-independent OU processes whose reversion speeds are generated according to $\pi$ (Fasen and Klüppelberg, 2007). Indeed, the large-population limit is justified in the sense of distributions (Yoshioka et al., 2023).

Importance of the condition (1) can be highlighted by the following theoretical results. Firstly, the stationarity of the supOU process (3) is satisfied under (1). Moreover, the stationary statistics are analytically obtained as follows with $M_k = \int_0^{+\infty} z^k v(\mathrm{d}z)$ ($k \in \mathbb{N}$) if it exists (Yoshioka, 2021): the mean

$$\mathbb{E}[X_t] = M_1 R, \tag{5}$$

variance

$$\mathbb{E}\left[(X_t - \mathbb{E}[X_t])^2\right] = \frac{M_2 R}{2}, \tag{6}$$

(unnormalized) skewness

$$\mathbb{E}\left[(X_t - \mathbb{E}[X_t])^3\right] = \frac{M_3 R}{3}, \tag{7}$$

and (unnormalized) kurtosis

$$\mathbb{E}\left[(X_t - \mathbb{E}[X_t])^4\right] - 3\mathbb{E}\left[(X_t - \mathbb{E}[X_t])^2\right] = \frac{M_4 R}{4}. \tag{8}$$

As clearly shown in (5)-(8), all these statistics are proportional to the inverse moment $R$ of (1). The accurate identification of $\pi$ is crucial in this view.

The autocorrelation function $\mathrm{ACF}(\tau)$ with time lag $\tau \geq 0$ can also be represented with $\pi$ (Barndorff-Nielsen, 2001):

$$\mathrm{ACF}(\tau) = \frac{1}{R} \int_0^{+\infty} \frac{\exp(-r\tau)}{r} \pi(\mathrm{d}r). \qquad (9)$$

The autocorrelation depends on both the inverse moment $R$ and some exponential integral with $\pi$, again highlighting the important role of $\pi$ that plays in the supOU process.

Hereafter, we always assume the Gamma-type reversion measure (2) as it is applicable to real time series data and gives the closed-form statistics. Indeed, we have $R = \{\beta(\alpha-1)\}^{-1}$ and the autocorrelation is explicitly obtained as the subexponential one

$$\mathrm{ACF}(\tau) = \left(\frac{1}{1+\beta\tau}\right)^{\alpha-1}, \quad \tau \geq 0. \qquad (10)$$

Then, the supOU process has a long memory if $\alpha \in (1,2]$. The autocorrelation is still far from being exponential for $\alpha > 2$ small. The exponential autocorrelation $\exp(-R^{-1}\tau)$ is recovered under the limit $\alpha \to 1$ with $R$ kept constant.

### 2.1.2 Model ambiguity

The model ambiguity is the distortion of the reversion measure through a Radon–Nikodym derivative $\phi$ as a $\pi$-a.s. positive measurable mapping between the reversion measure $\pi(\mathrm{d}r)$ and a distorted one $\pi_\phi(\mathrm{d}r)$ as an absolutely continuous measure with respect to $\pi$, where $\pi_\phi(\mathrm{d}r) = \phi(r)\pi(\mathrm{d}r)$ and

$$\int_0^{+\infty} \pi_\phi(\mathrm{d}r) = \int_0^{+\infty} \phi(r)\pi(\mathrm{d}r) = 1. \qquad (11)$$

We consider the ambiguity in the sense of Tsallis (e.g., Zou et al. (2022)), which means that the inverse moment $R$ should be effectively replaced by the following distorted one

$$R_{\phi,q} = \int_0^{+\infty} \frac{1}{r} \{\phi(r)\}^q \pi(\mathrm{d}r) \qquad (12)$$

with the shape parameter $q > 0$. The corresponding statistics under the model ambiguity can be obtained by the formal replacement $R \to R_{\phi,q}$ in (5)-(8). The most natural choice would be the case $q = 1$, the classical EVaR case, while we will show that this choice does not work for the entropic evaluation of the statistics of interest.

***Remark 1.*** We focus on the jump-driven supOU processes that do not have Brownian diffusive noises because it is suitable for our application. Nevertheless, the exclusion of the Brownian noise does not significantly limit the applicability of the proposed methodology because the inverse moment $R$ also appears in the variance in the diffusive case as well (Barndorff-Nielsen, 2001).

We end this section by recalling *q*-exponential and *q*-logarithm functions that will be frequently used in this paper (e.g., Ma and Tian (2021)):

$$\exp_q(x) = \begin{cases} (1+(1-q)x)^{\frac{1}{1-q}} & (q \neq 1) \\ \exp(x) & (q = 1) \end{cases}, \quad 1+(1-q)x > 0 \tag{13}$$

and

$$\ln_q(x) = \begin{cases} \dfrac{x^{1-q}-1}{1-q} & (q \neq 1) \\ \ln(x) & (q = 1) \end{cases}, \quad x > 0. \tag{14}$$

The functions $\exp_q$ and $\ln_q(x)$ are increasing, and are strongly convex and concave in their domains, respectively, and reduce to the classical exp and ln functions if $q = 1$ (**Figures 1** and **2**).

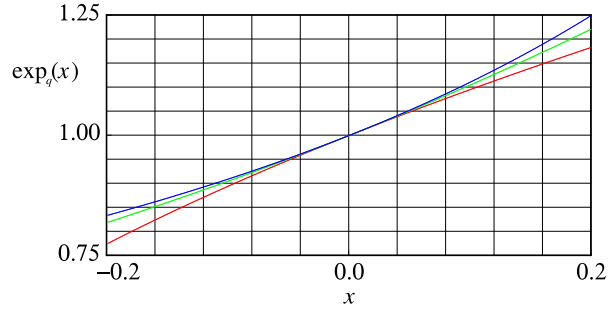

**Figure 1.** Profiles of the $\exp_q$ functions: $q = 0.5$ (Red), $q = 1.0$ (Green), and $q = 1.5$ (Blue).

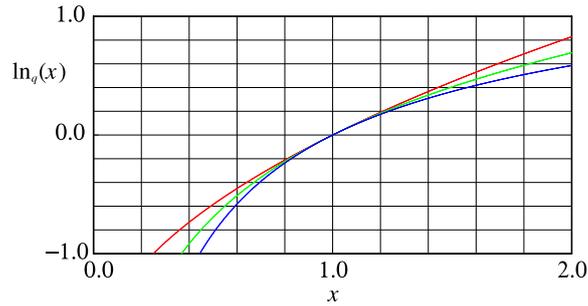

**Figure 2.** Profiles of $\ln_q$ functions: $q = -1$ (Red), $q = 1$ (Green), and $q = 2$ (Blue).

## 2.2 Tsallis Value-at-Risk
### 2.2.1 Upper Tsallis Value-at-Risk

The upper-Tsallis Value-at-Risk (upper-TsVaR) in this paper is the TsVaR of Zou et al. (2022) to give an upper bound of the expectation of a random variable by using a convex risk measure. We define TsVaRs with an adaptation to our setting. The upper-TsVaR with the accuracy parameter $a \in (0,1]$, denoted as $\overline{\text{TsVaR}}_a$, to give a worst-case overestimation of the inverse moment $R$ is defined as follows:

$$\overline{\text{TsVaR}}_a = \inf_{\lambda>0}\left\{\frac{1}{\lambda}\ln_q\left(\frac{1}{a}\int_0^{+\infty}\exp_q\left(\frac{\lambda}{r}\right)\pi(\mathrm{d}r)\right)\right\} \left(\equiv \inf_{\lambda>0}\{\overline{g}(\lambda)\}\right). \quad (15)$$

The function $\overline{g}(\lambda)$ is motivated from the Chernoff inequality (Zou et al., 2022) of a sharp upper bound of the probability $\Pr\left(\frac{1}{r} \geq g\right)$ under the probability measure $\pi$, namely the largest solution to the right equation below:

$$\Pr\left(\frac{1}{r} \geq g\right) \leq \frac{\int_0^{+\infty}\exp_q\left(\frac{\lambda}{r}\right)\pi(\mathrm{d}r)}{\exp_q(\lambda g)} = a \quad \text{for all} \quad \lambda > 0. \quad (16)$$

This $\overline{g}(\lambda)$ serves as an upper bound of the Value-at-Risk with the quantile $a$ called the accuracy parameter, thereby the upper-TsVaR as its lower bound. Namely, the upper-TsVaR is $\overline{g}(\lambda)$ closed to the corresponding Value-at-Risk. The conventional (upper-)EVaR is recovered if $q = 1$ in (15) (Ahmadi-Javid, 2012). Note that the upper-TsVaR (15) can be rewritten as

$$\overline{\text{TsVaR}}_a = \inf_{\lambda>0}\left\{\lambda \ln_q\left(\frac{1}{a}\int_0^{+\infty}\exp_q\left(\frac{1}{\lambda r}\right)\pi(\mathrm{d}r)\right)\right\} \left(= \inf_{\lambda>0}\{\overline{g}(\lambda^{-1})\}\right). \quad (17)$$

We use (15) and (17) interchangeably depending on the context.

We need to suitably choose the shape parameter $q$ so that the following integrability condition is satisfied:

$$\int_0^{+\infty}\exp_q\left(\frac{\lambda_0}{r}\right)\pi(\mathrm{d}r) < +\infty \quad \text{with some} \quad \lambda_0 > 0. \quad (18)$$

**Proposition 1** below shows that the condition (18) is not satisfied if $q$ is large. Hence, the classical EVaR cannot be used in our case.

*Proposition 1*

*The condition (18) is satisfied if and only if $0 < q < 1 - \alpha^{-1}$.*

Considering **Proposition 1**, we limit our focus to the parameter range $0 < q < 1 - \alpha^{-1}$ for the upper-TsVaR. **Proposition 2** is of importance in the application of the upper-TsVaR, especially for its computation.

*Proposition 2*

*The upper-TsVaR is a convex risk measure; especially, $\overline{g}(\lambda)$ is convex at each $\lambda > 0$. Furthermore, the infimum of (17) is attained in a compact set $C$ of the form $[0, \overline{C}]$ with a constant $\overline{C} > 0$.*

### 2.2.2 Lower Tsallis Value-at-Risk

The lower-Tsallis Value-at-Risk (upper-TsVaR) is an under-estimation counterpart of the upper-TsVaR to evaluate a lower bound of the expectation of a random variable. The lower-TsVaR with the accuracy parameter $a \in (0,1]$, denoted as $\underline{\text{TsVaR}}_a$, to give a worst-case overestimation of the inverse moment $R$ is defined as follows:

$$\underline{\text{TsVaR}}_a = \sup_{\lambda>0}\left\{-\frac{1}{\lambda}\ln_q\left(\frac{1}{a}\int_0^{+\infty}\exp_q\left(-\frac{\lambda}{r}\right)\pi(\mathrm{d}r)\right)\right\} \left(\equiv \sup_{\lambda>0}\{\underline{g}(\lambda)\}\right). \tag{19}$$

We also use the following equivalent formulation interchangeably depending on the context.

$$\underline{\text{TsVaR}}_a = \sup_{\lambda>0}\left\{-\lambda\ln_q\left(\frac{1}{a}\int_0^{+\infty}\exp_q\left(-\frac{1}{\lambda r}\right)\pi(\mathrm{d}r)\right)\right\} \left(= \sup_{\lambda>0}\{\underline{g}(\lambda^{-1})\}\right). \tag{20}$$

The lower-TsVaR is based on the another Chernoff inequality:

$$\Pr\left(\frac{1}{r} \leq g\right) \leq \frac{\int_0^{+\infty}\exp_q\left(-\frac{\lambda}{r}\right)\pi(\mathrm{d}r)}{\exp_q(-\lambda g)} = a \quad \text{for all} \quad \lambda > 0. \tag{21}$$

Indeed, by the classical Markov inequality combined with the increasing property of $\exp_q$, we have

$$\Pr\left(\frac{1}{r} \leq g\right) = \Pr\left(\exp_q\left(-\frac{\lambda}{r}\right) \geq \exp_q(-\lambda g)\right) \leq \frac{\int_0^{+\infty}\exp_q\left(-\frac{\lambda}{r}\right)\pi(\mathrm{d}r)}{\exp_q(-\lambda g)}, \quad \lambda > 0. \tag{22}$$

The largest solution to the right equation of (21) is $g = \underline{g}$. This $\underline{g}$ serves as a lower bound of the Value-at-Risk with the quantile $a$, thereby the lower-TsVaR as its upper bound. Namely, the upper-TsVaR is $\underline{g}(\lambda)$ closest to the Value-at-Risk.

As in the case of the upper-TsVaR, the range of the shape parameter $q$ must be limited to well-define the lower-TsVaR. Namely, we need

$$\int_0^{+\infty}\exp_q\left(-\frac{\lambda_0}{r}\right)\pi(\mathrm{d}r) < +\infty \quad \text{with some} \quad \lambda_0 > 0 \tag{23}$$

to well-define (19). **Proposition 3** below shows that the condition (23) is not satisfied if $q$ is small. In this case, the classical EVaR ($q=1$) is allowed.

*Proposition 3*

*The condition (23) is satisfied if and only if $q \geq 1$.*

As in the upper-TsVaR case, the convexity and compactness results follow for the lower-TsVaR. The proof of the **Proposition 4** below is omitted as it is essentially the same with that of **Proposition 2**.

*Proposition 4*

*The lower-TsVaR is a convex risk measure; especially, $\underline{g}(\lambda)$ is convex at each $\lambda > 0$. Furthermore, the infimum of (20) is attained in a compact set $C$ of the form $[0, \underline{C}]$ with a constant $\underline{C} > 0$.*

We state **Proposition 5** below showing that the lower- and upper-TsVaRs are indeed lower and upper bounds of the inverse moment $R$, respectively. **Proposition 5** also suggests that calling the parameter $a$ appearing in the TsVaRs as the accuracy parameter is reasonable because the TsVaRs get closer to $R$ as $a$ increases.

*Proposition 5*

*The following inequalities hold true for any $0 < a \le b \le 1$:*

$$\overline{\text{TsVaR}}_a \ge \overline{\text{TsVaR}}_b \ge \overline{\text{TsVaR}}_1 \ge R \tag{24}$$

*for $0 < q < 1 - \alpha^{-1}$ and*

$$\underline{\text{TsVaR}}_a \le \underline{\text{TsVaR}}_b \le \underline{\text{TsVaR}}_1 \le R \tag{25}$$

*for $q \ge 1$.*

□

***Remark 2.*** The notion of lower-TsVaR can be applied to generic non-negative random variables under a suitable assumption of its integrability with respect to a provability measure.

## 2.3 Duality

A duality result between the TsVaRs and corresponding nonlinear expectations are provided, which serves as a pathway to obtain the worst-case distorted reversion measures. For the upper-TsVaR with $0 < q < 1 - \alpha^{-1}$, the following relationship holds true by the direct application of Lemmas 3.6 and 3.7, and Theorem 3.8 of Zou et al. (2022) to our setting:

$$\overline{\text{TsVaR}}_a = \sup_{\phi \in \mathfrak{F}} \left\{ a^{q-1} \int_0^{+\infty} \{\phi(r)\}^q \frac{1}{r} \pi(\mathrm{d}r) \right\} \tag{26}$$

with the admissible set $\mathfrak{F}$ of Radon–Nikodym derivatives

$$\mathfrak{F} = \left\{ \phi : (0, +\infty) \to (0, +\infty) \Big| \int_0^{+\infty} \phi(r) \pi(\mathrm{d}r) = 1, \ H_q(\phi) \le -\ln_q(a) \right\} \tag{27}$$

and the Tsallis divergence

$$H_q(\phi) = \frac{1}{1-q} \left( 1 - \int_0^{+\infty} \{\phi(r)\}^q \pi(\mathrm{d}r) \right), \tag{28}$$

and further

$$\ln_q \left( \int_0^{+\infty} \exp_q\left(\frac{\lambda}{r}\right) \pi(\mathrm{d}r) \right) = \sup_{\phi \in \mathfrak{F}'} \left\{ \lambda \int_0^{+\infty} \{\phi(r)\}^q \frac{1}{r} \pi(\mathrm{d}r) - H_q(\phi) \right\} \tag{29}$$

with

$$\mathfrak{F}' = \left\{ \phi : (0, +\infty) \to (0, +\infty) \middle| \int_0^{+\infty} \phi(r) \pi(\mathrm{d}r) = 1 \right\}. \quad (30)$$

These results become useful when computing the

We obtain a similar duality relationship like (26) and (29) for the lower-TsVaR.

*Proposition 6*

*Assume $q \geq 1$. Then, it follows that*

$$-\ln_q \left( \int_0^{+\infty} \exp_q \left( \frac{-\lambda}{r} \right) \pi(\mathrm{d}r) \right) = \inf_{\phi \in \mathfrak{F}'} \left\{ \lambda \int_0^{+\infty} \{\phi(r)\}^q \frac{1}{r} \pi(\mathrm{d}r) + H_q(\phi) \right\} \quad (31)$$

*and further*

$$\underline{\mathrm{TsVaR}}_a = \inf_{\phi \in \mathfrak{F}} \left\{ a^{q-1} \int_0^{+\infty} \{\phi(r)\}^q \frac{1}{r} \pi(\mathrm{d}r) \right\}. \quad (32)$$

## 2.4 Computational method

### 2.4.1 TsVaRs

We propose a simple gradient descent method to compute the TsVaRs. The gradient descent method is a temporal discretization for computing an equilibrium point of the ordinary differential equations (ODEs):

$$\frac{\mathrm{d}\lambda}{\mathrm{d}u} = -\frac{\partial \overline{g}(\lambda^{-1})}{\partial \lambda} \quad (33)$$

for the upper-TsVaR (minimization problem) and

$$\frac{\mathrm{d}\lambda}{\mathrm{d}u} = \frac{\partial \underline{g}(\lambda^{-1})}{\partial \lambda} \quad (34)$$

for the lower-TsVaR (maximization problem) where $\lambda$ here is a time-dependent curve $\lambda = \lambda(u)$ parameterized by a pseudo time $u \geq 0$ with some initial condition $\lambda(0)$ (Fixed to 500 in this paper but other positive values can be used as explained below). We use the representations (17) and (20) with which the equilibrium point of each ODE exists in a compact set by **Propositions 1** and **3**. These ODEs can be integrated from the initial time 0 until the solutions approach sufficiently close to the equilibrium. A TsVaR can then be obtained by substituting the equilibrium value to $\overline{g}$ or $\underline{g}$.

Other than the temporal discretization, the full implementation of the gradient descent requires to numerically evaluate the integrals with respect to the reversion measure $\pi$ as they will not be available in closed-forms. We use the quantile discretization method of Yoshioka et al. (2023) to discretize the reversion measure of the gamma type (2) and analyze its convergence computationally. We do not directly use their method for the upper-TsVaR because it discretizes the reversion measure $\pi$ into the discrete empirical one, denoted as $\pi_N = \frac{1}{N} \sum_{i=1}^{N} \delta(r - r_i)$, and performs the numerical integration

$$\int_0^{+\infty} \exp_q\left(\frac{\lambda}{r}\right)\pi(\mathrm{d}r) \to \frac{1}{N}\sum_{i=1}^{N} \exp_q\left(\frac{\lambda}{r_i}\right). \tag{35}$$

Here, for $1 \leq i \leq n$, $r_i$ is the quantile satisfying $\frac{2i-1}{2N} = \int_0^{r_i} \pi(\mathrm{d}r)$, $\delta(r-r_i)$ is the Dirac Delta at $r = r_i$. The integrand $\exp_q\left(\frac{\lambda}{r}\right)$ blows up at the speed $r^{\frac{-1}{1-q}}$, which has preliminary been found to significantly degrade the accuracy of the numerical integration (35). We mitigate this issue by exploiting the following equality

$$\int_0^{+\infty} \exp_q\left(\frac{\lambda}{r}\right)\pi(\mathrm{d}r) = \frac{\Gamma(\alpha - 1/(1-q))}{\Gamma(\alpha)} \beta^{-1/(1-q)} \int_0^{+\infty} r^{\frac{1}{1-q}} \exp_q\left(\frac{\lambda}{r}\right) \tilde{\pi}(\mathrm{d}r) \tag{36}$$

with the new Gamma density (recall **Proposition 1**)

$$\tilde{\pi}(\mathrm{d}r) = \frac{1}{\Gamma(\alpha - 1/(1-q))\beta^{\alpha-1/(1-q)}} r^{\alpha - \frac{1}{1-q} - 1} \exp\left(-\frac{r}{\beta}\right)\mathrm{d}r. \tag{37}$$

The integrand in the right-hand side of (36) is now bounded for small $r > 0$, and hence the aforementioned singularity issue is completely resolved. The discretization is then carried out for $\tilde{\pi}$. Namely, for each fixed $q$, we perform the numerical integration

$$\int_0^{+\infty} r^{\frac{1}{1-q}} \exp_q\left(\frac{\lambda}{r}\right)\tilde{\pi}(\mathrm{d}r) \to \frac{1}{N}\sum_{i=1}^{N} r_i^{\frac{1}{1-q}} \exp_q\left(\frac{\lambda}{r_i}\right) \tag{38}$$

with each $r_i$ the quantile satisfying $\frac{2i-1}{2N} = \int_0^{r_i} \tilde{\pi}(\mathrm{d}r)$ ($1 \leq i \leq N$). The aforementioned singularity issue is not encountered in computing the lower-TsVaR, thereby we use (35) for its computation.

The temporal discretization that we use is a semi-implicit one. More specifically, (33) is rewritten as

$$\begin{aligned}\frac{\mathrm{d}\lambda}{\mathrm{d}u} &= -\frac{\partial}{\partial \lambda}\left\{\lambda \ln_q\left(\frac{1}{a}\int_0^{+\infty} \exp_q\left(\frac{1}{\lambda r}\right)\pi(\mathrm{d}r)\right)\right\} \\ &= -\lambda\frac{1}{\lambda}\ln_q\left(\frac{1}{a}\int_0^{+\infty}\exp_q\left(\frac{1}{\lambda r}\right)r^{\frac{1}{1-q}}\tilde{\pi}(\mathrm{d}r)\right)(\equiv -I_1) \\ &\quad + \frac{1}{a\lambda}\int_0^{+\infty}\frac{1}{r}\left\{\exp_q\left(\frac{1}{\lambda r}\right)\right\}^q r^{\frac{1}{1-q}}\tilde{\pi}(\mathrm{d}r)\left(\frac{1}{a}\int_0^{+\infty}\exp_q\left(\frac{1}{\lambda r}\right)r^{\frac{1}{1-q}}\tilde{\pi}(\mathrm{d}r)\right)^{-q}(\equiv I_2)\end{aligned} \tag{39}$$

If $\lambda > 0$, then $I_1, I_2 > 0$. These fixed signs motivate a semi-implicit discretization that handle $I_1$ implicitly while $I_2$ explicitly:

$$\frac{\lambda^{(n+1)} - \lambda^{(n)}}{\Delta u} = -\frac{\lambda^{(n+1)}}{\lambda^{(n)}}I_1^{(n)} + I_2^{(n)} \quad \text{or} \quad \lambda^{(n+1)} = \left(1 + \frac{1}{\lambda^{(n)}}I_1^{(n)}\Delta u\right)^{-1}\left(I_2^{(n)}\Delta u + \lambda^{(n)}\right) \tag{40}$$

with the superscript $(n)$ representing the quantity evaluated at the $n$th iteration and $\Delta u > 0$ the step size. The denominator of (40) is positive as long as $\lambda^{(n)} > 0$, hence an induction method yields $\lambda^{(n)} > 0$ ($n = 1, 2, 3, ...$) if $\lambda^{(0)} > 0$. The semi-implicit discretization therefore preserves the positivity of $\lambda$ irrespective to $\Delta u$, even if it is very large. The same discretization is applied to computing the lower-TsVaR based on the integration using $\pi$. We set $\Delta u = 10^4$ in this paper.

***Remark 3*** The equilibrium of the gradient descent, if it is attained, is unique owing to the strict convexity shown in **Appendix A.2**.

### 2.4.2 Radon–Nikodym derivatives

The duality results presented in the previous subsection is exploited to numerically compute the worst-case Radon–Nikodym derivatives realizing the TsVaRs. This point has not been explored in detail in Zou et al. (2022). We explain the methodology to compute the worst-case Radon–Nikodym derivative for the lower-TsVaR, but the same procedure applies to the upper-TsVaR due to their symmetric representations as implied so far in this paper. The pseud-algorithm is given as follows (**Pseudo-Algorithm 1**).

*Pseudo-Algorithm 1*

**Step 0.** *Set $\pi$ and the shape parameter $q$.*

**Step 1.** *Set some $\lambda = \lambda_0$ in (31).*

**Step 2.** *Solve the optimization problem in the right-hand side of (31) and obtain the minimizing Radon–Nikodym derivative $\phi = \phi_*(r)$ that is parameterized by $\lambda_0$.*

**Step 3.** *Compute the corresponding accuracy parameter $a = a_*$ from the equality*

$$H_q(\phi_*) = -\ln_q(a_*) \qquad (41)$$

*as $a_* = \exp_q(-H_q(\phi_*))$.*

**Step 4.** *Compute the maximizer $\lambda = \lambda_*$ in the right-hand side of (19) by using the gradient descent method.*

**Step 5.** *Find the lower-TsVaR by substituting $\lambda = \lambda_*$ to the right-hand side of (19).*

The key point in **Pseudo-Algorithm 1** is that its main part starts from finding a Radon–Nikodym derivative related to a nonlinear expectation, the corresponding accuracy parameter, and finally the corresponding TsVaR. In this view, the Radon–Nikodym derivative and the TsVaR are linked with each other through the equality (41). This equality comes from the boundary of the admissible set of the decision variables (27). **Step 3** combined with **Remark 3** suggests that the optimization in (31) is achieved at a unique point.

**Pseudo-Algorithm 1** can be used to extensively analyze the relations among the Radon–Nikodym derivative, accuracy parameter, and TsVaR by examining many values of $\lambda = \lambda_0$ in **Step 1**. We employ this strategy in **Section 3**. From a practical viewpoint, this strategy corresponds to obtaining the risk measures for diverse values of the ambiguity-aversion of the environmental mangers because the parameter $\lambda = \lambda_0$ in **Step 1** can be understood as an ambiguity-aversion parameter whose larger value corresponds to the manager who is more ambiguity averse.

***Remark 4.*** The minimizer of (32) is unique due to the strict convexity, which is the worst-case Radon–Nikodym derivative obtained at **Step 2**. Therefore, given the worst-case Radon–Nikodym derivative, the corresponding TsVaR is determined uniquely.

## 3. Application

### 3.1 Study area

The study area is Tedori River in Ishikawa Prefecture, Japan (**Figure 3**). The main channel of Tedori River starts in the Hakusan Mountains at an elevation of 2702 m. The river channel has a length of 72 km and a basin area of 807 km$^2$, and it pours into the Sea of Japan (Iwasaki-Yoshioka et al., 2016). The alluvial fan of the river has an area of 170 km$^2$, which is covered by paddy fields. Thus, the hydrology of Tedori River is important for assessing the paddy, surface water, and groundwater environments in the study area (Kimura et al., 2022; Takase and Fujihara, 2019). Studies on the impacts of extensive human intervention to Tedori River have focused on the riparian vegetation and river morphology, and they have shown that dams have modulated the sediment budget and hence the riverbed dynamics (Dang et al., 2014; Nallaperuma and Asaeda, 2019). These results imply that the streamflow discharge is crucial for evaluating the environment, ecosystem, and risk to human life within the study area. However, a detailed time series analysis has not been performed here to the best of our knowledge, which motivated us to apply our proposed methodology to the study area.

There are three observation stations along the main channel of Tedori River: Tsurugi (14.30 km from the sea), Nakajima (19.00 km from the sea), and Kazarashi (52.20 km from the sea). Hourly records of the streamflow discharge were obtained from 2018 January 1 01:00:00 JST to 2021 December 31 23:50:00 JST (Water Information System by Ministry of Land, Infrastructure, Transport and Tourism, Japan; http://www1.river.go.jp/) (**Figure 4**). Tsurugi is almost at the apex of the alluvial fan, so it is considered to govern the surface water and groundwater dynamics in the fan. There is an agricultural water intake between Tsurugi and Nakajima, which implies that the streamflow discharge at Tsurugi is influenced by agricultural water use. There are a few dams between Nakajima and Kazarashi while there is no dam upstream of Kazarashi. Therefore, Kazarashi receives the least human intervention while Nakajima is affected by dam-induced flow alteration. From winter to spring, snowmelt contributes to the streamflow discharge. Therefore, each of the three observation stations is important for understanding the streamflow dynamics of Tedori River.

We assumed tempered stable or gamma subordinators as the background jump processes:

$$v(\mathrm{d}z) = A\frac{\exp(-Bz)}{z^{1+C}}\mathrm{d}z, \quad z > 0 \qquad (42)$$

where $A > 0$, $B > 0$, and $C < 1$. This obtains $M_k = AB^{C-k}\Gamma(k-C) < +\infty$ ($k \in \mathbb{N}$). Therefore, the identification method of Yoshioka et al. (2023) can be applied, as explained below. First, we identified the reversion measure $\pi$ (i.e., the parameters $\alpha, \beta$) by nonlinear least-squares fitting between the empirical and theoretical autocorrelation functions. In each case, the fitting is carried out only for the largest time lag where the observed autocorrelation function remained positive. The minimum discharge $\underline{X} \geq 0$ during the study period is identified from the observation data. After we identified the reversion measure $\pi$ and minimum discharge $\underline{X}$, the parameters $A, B, C$ and the minimum discharge are identified by another nonlinear least-squares fitting to minimize the sum of the squared relative errors of the average, variance, skewness, and kurtosis of the shifted process $X + \underline{X}$ (i.e., the discharge at an observation station). Note that the constant shift only affects the average among the analyzed statistics.

**Table 1** summarizes the empirical and computed statistics as well as the identified parameter values. Fitted parameter values at each observation station are summarized in **Table 2**. The maximum relative errors are at most 2 to 3 % at each station. **Figure 5** shows the empirical and fitted theoretical autocorrelation functions at each observation station. The parameter $\alpha$ is around 1.7 and hence smaller than 1 at all stations, which indicates that the streamflow discharges at these stations have long memories. Interestingly, the identification results for the Lévy measure suggest that the jumps at Tsurugi and Nakajima were driven by processes with infinite activity (i.e., infinitely many small jumps existed in each bounded open time interval in the sense that $\int_0^{+\infty} v(\mathrm{d}z) = +\infty$). In contrast, the discharge at Kazarashi (i.e., the most upstream station) is driven by compound Poisson jumps (i.e., finitely many small jumps existed in each bounded open time interval in the sense that $\int_0^{+\infty} v(\mathrm{d}z) < +\infty$). One reason for this qualitative difference is the presence of dams downstream of Kazarashi, which artificially modulated the discharge flow. **Table 2** shows that the long-memory nature is strongest at Kazarashi, which has the smallest $\alpha$, followed by Nakajima and Tsurugi in order. The weakening of the long-memory structure from upstream to downstream can be attributed to human intervention in the study area (e.g., dam operation and agricultural water intake) in the study area. Clarifying the relationship between the memory structure and anthropogenic impacts is beyond the scope of this paper, but it is an important future topic that will be addressed elsewhere.

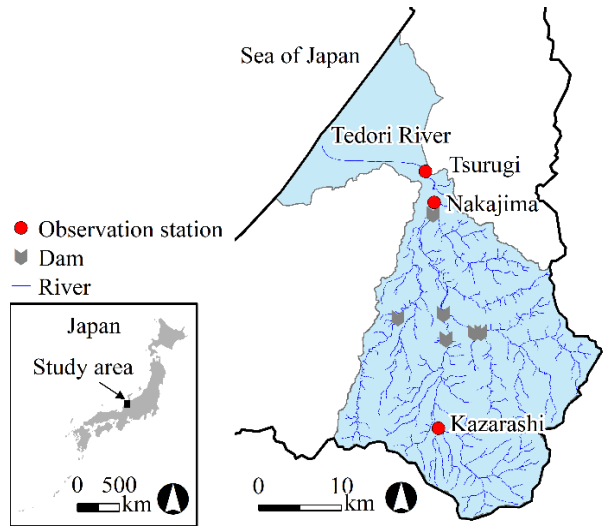

**Figure 3.** Map of the study area.

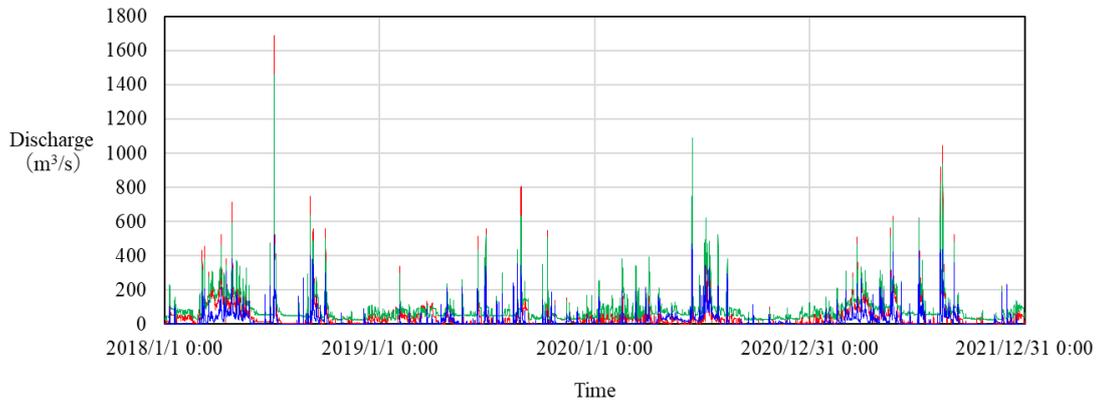

**Figure 4.** Hourly discharge data at the three observation stations: Tsurugi (red), Nakajima (green), and Kazarashi (blue).

**Table 1.** Statistics at the study sites of Cases 1 and 2.

| Statistics | Tsurugi | | Nakajima | | Kazarashi | |
|---|---|---|---|---|---|---|
| | Observed | Fitted | Observed | Fitted | Observed | Fitted |
| Average (m$^3$/s) | 4.673.E+01 | 4.692.E+01 | 8.502.E+01 | 8.523.E+01 | 2.042.E+01 | 2.074.E+01 |
| Variance (m$^6$/s$^3$) | 5.850.E+03 | 5.838.E+03 | 5.327.E+03 | 5.321.E+03 | 1.624.E+03 | 1.615.E+03 |
| Skewness (-) | 4.842.E+00 | 5.000.E+00 | 3.631.E+00 | 3.714.E+00 | 3.788.E+00 | 3.762.E+00 |
| Kurtosis (-) | 4.470.E+01 | 4.403.E+01 | 2.591.E+01 | 2.564.E+01 | 2.049.E+01 | 2.073.E+01 |

**Table 2.** Statistics at the study sites of Cases 1 and 2.

| Parameter | Tsurugi | Nakajima | Kazarashi |
|---|---|---|---|
| $\underline{X}$ (m$^3$/s) | 0.07 | 9.92 | 0.34 |
| $\alpha$ (-) | 1.73 | 1.76 | 1.67 |
| $\beta$ (1/hour) | 0.0372 | 0.0219 | 0.0544 |
| $A$ (hour·s$^{-C}$·m$^{3C}$) | 0.0125 | 0.0382 | 0.0000358 |
| $B$ (s/m$^3$) | 0.00309 | 0.00378 | 0.0146 |
| $C$ (-) | 0.230 | 0.467 | -1.31 |

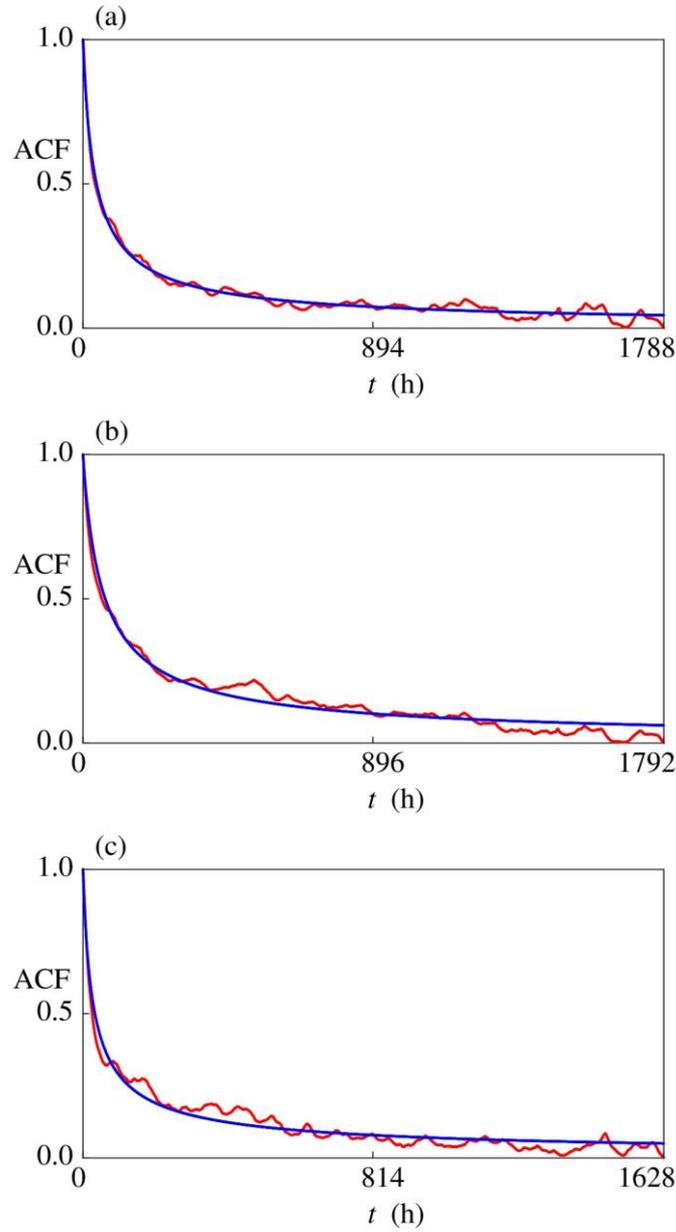

**Figure 5.** Empirical (red) and fitted theoretical (blue) autocorrelation functions (ACFs) at (a) Tsurugi, (b) Nakajima, and (c) Kazarashi

## 3.2 Computational results

### 3.2.1 Convergence analysis

Here, we present a computational example of the worst-case Radon–Nikodym derivative and TsVaRs based on **Pseudo-Algorithm 1**, along with the convergence analysis. The error sources of the computation are the discretization error of the reversion measure $\pi$ and the truncation error of the gradient descent. In the current setting, the latter is significantly smaller than the former because the former is terminated when the successive difference in $\lambda$ became less than $10^{-12}$. The degrees of freedom to discretize the reversion measure are set to $N = 2^m$ with $m \in \mathbb{N}$ following Yoshioka et al. (2023). Analytical solutions to the TsVaR have not been found, so we are unable to fully evaluate the computational results. Instead, the computational result at $N = 2^{17} = 131,072$ is regarded as a reference solution for evaluating results with coarser resolutions. The accuracy parameter $a$ is set to 0.99 or 0.60 depending on the analysis, as explained below. Only the parameter set for Kazarashi is used here because the conclusions hold qualitatively true for the remaining stations.

**Tables 3** and **4** present the computational results, which suggest that using a finer discretization led to a smaller error with respect to the reference solution. In particular, the convergence of the lower-TsVaR is satisfactory at $m = 12$ for both $a = 0.99$ and $a = 0.60$. In addition, when the upper-TsVaR is being computed, the convergence of the numerical integration is slower with (35) than with (38). This indicates that (35) is suitable. Based on the obtained computational results, we use the resolution $N = 2^{15} = 32,768$ for all following computations.

**Table 3.** Convergence of the TsVaRs for different values of $N = 2^m$ ($a = 0.99$).

|       | Upper-TsVaR: (35) |        | Upper-TsVaR: (38) |        | Lower-TsVaR |         |
|-------|-------------------|--------|-------------------|--------|-------------|---------|
| $M$   | Computed          | Error  | Computed          | Error  | Computed    | Error   |
| 12    | 32.2283           | 1.9304 | 37.2471           | 0.6679 | 19.8581     | 0.00000 |
| 13    | 32.6964           | 1.4623 | 37.5876           | 0.3274 | 19.8581     | 0.00000 |
| 14    | 33.1174           | 1.0413 | 37.7587           | 0.1563 | 19.8581     | 0.00000 |
| 15    | 33.4980           | 0.6607 | 37.8463           | 0.0687 | 19.8581     | 0.00000 |
| 16    | 33.8436           | 0.3151 | 37.8915           | 0.0235 | 19.8581     | 0.00000 |
| 17    | 34.1587           |        | 37.9150           |        | 19.8581     |         |

**Table 4.** Convergence of the TsVaRs for different values of $N = 2^m$ ($a = 0.60$).

|       | Upper-TsVaR: (35) |        | Upper-TsVaR: (38) |        | Lower-TsVaR |         |
|-------|-------------------|--------|-------------------|--------|-------------|---------|
| $m$   | Computed          | Error  | Computed          | Error  | Computed    | Error   |
| 12    | 61.3396           | 7.2828 | 83.6377           | 0.0103 | 7.58625     | 0.00008 |
| 13    | 63.0637           | 5.5587 | 83.6428           | 0.0052 | 7.58620     | 0.00003 |
| 14    | 64.6396           | 3.9828 | 83.6455           | 0.0025 | 7.58618     | 0.00001 |
| 15    | 66.0828           | 2.5396 | 83.6469           | 0.0011 | 7.58617     | 0.00000 |
| 16    | 67.4065           | 1.2159 | 83.6476           | 0.0004 | 7.58617     | 0.00000 |
| 17    | 68.6224           |        | 83.6480           |        | 7.58617     |         |

### 3.2.2 Computation for all the observation stations

The Radon–Nikodym derivative and corresponding lower- and upper-TsVaRs are computed for the three observation stations at the resolution $N = 2^m$, which the above convergence analysis suggested is sufficiently accurate for our purposes. Unless otherwise specified, we set $q = 0.33$ and $q = 1.33$ for the upper- and lower-TsVaRs, respectively. Recall that the allowable range of $q$ is restricted to $(0, 1 - \alpha^{-1})$ for the upper-TsVaR. The upper bounds $1 - \alpha^{-1}$ at Tsurugi, Nakajima, and Kazarashi are 0.42, 0.43, and 0.40, respectively. Hereafter, we focus on the normalized TsVaR because it quantifies the relative error of the first- to fourth-order statistics in a unified way according to (5)–(8). The normalized TsVaR is given

by

$$\text{Rat} = \frac{\overline{\text{TsVaR}}}{R} \text{ or } \frac{\underline{\text{TsVaR}}}{R}. \tag{43}$$

For each computational case, we use **Pseudo-Algorithm 1** for which the accuracy parameter $a = a^*$ of **Step 3** is set from 0.999 to some numerically obtained lower bound depending on the computational condition (around 0.6 for upper-TsVaRs and around 0.2 for lower-TsVaRs).

**Figures 6** and **7** show the computed upper- and lower-TsVaRs, respectively, at each station. **Figure 6** shows that, among the three stations, Kazarashi has the smallest upper-TsVaR, followed by Tsurugi and Nakajima in order. This can be attributed to the smaller $\alpha$ and hence longer memory at Kazarashi. In the supOU process, according to the reversion measure (2) and autocorrelation function (10), a longer memory corresponds to a stronger singularity of the reversion measure near the origin $r = 0$, which results in a higher sensitivity of $R$ to model ambiguity. This is because the behavior of the reversion measure near $r = 0$ becomes dominant as $\alpha$ decreases. The normalized lower-TsVaRs in **Figure 7** support this theoretical consideration because they are again smallest at Kazarashi, followed by Tsurugi and Nakajima in order (**Table 2**). However, the normalized lower-TsVaRs are not critically different from each other. These results show that the memory structure strongly affects the TsVaRs and thus the key statistical moments of the streamflow discharge.

**Figures 6** and **7** suggest that the accuracy parameter $a$ should not be set too small because this would cause the normalized TsVaRs to deviate significantly from 1, which would give an unrealistic estimate. However, $a$ is a user-specific parameter, so it should be determined by decision-makers such as environmental managers. In addition, the TsVaRs depend sharply and possibly superlinearly on the accuracy parameter near $a = 1$. For example, at Nakajima, the normalized upper-TsVaR is 1.16 at $a = 0.995$, 1.20 at $a = 0.990$, and 1.32 at $a = 0.985$. This parameter sensitivity should be considered in each application.

The rest of the analysis focuses on Kazarashi, which the previous computational results indicated is most susceptible among the stations to model ambiguity. **Figures 8** and **9** show the worst-case Radon–Nikodym derivatives for the overestimation and underestimation cases, respectively. Decreasing the accuracy parameter $a$ causes the Radon–Nikodym derivatives to become more sharply concentrated at the slower (i.e., overestimation case) or faster reversion speed (i.e., underestimation case). This suggests that decreasing the accuracy parameter, which corresponds to a decrease in confidence in the identified model, results in more strongly distorted reversion measure profiles. Thus, our proposed methodology can relate the model confidence to the Radon–Nikodym derivatives.

Finally, we analyze the $q$-dependence of the normalized TsVaRs at Kazarashi. **Figure 10** shows the normalized upper-TsVaR for $q = 0.33$ and $q = 0.37$. Similarly, **Figure 11** shows the normalized lower-TsVaR for $q = 1.33$ and $q = 2.00$. **Figure 10** suggests that choosing the larger $q$ led to a pessimistic estimate of the normalized upper-TsVaRs because of the stronger concave–convex nature of their functional form. **Figure 11** implies that the same applies to the normalized lower-TsVaRs. Hence, for a statistical evaluation of the streamflow discharge, the parameter $q$ can be utilized as a user-specific parameter that reflects their aversion to ambiguity; i.e., a more ambiguity-averse decision-maker should choose a larger value of $q$.

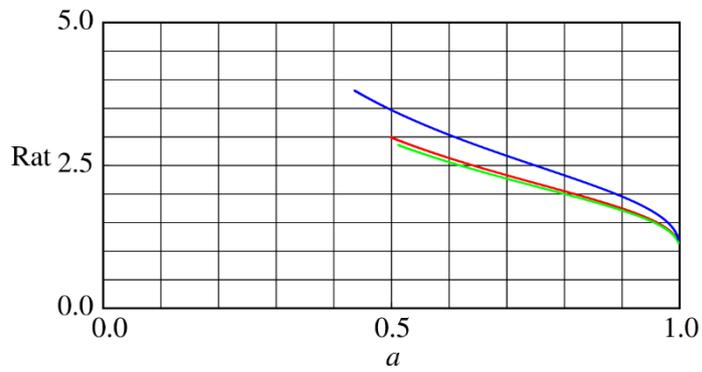

**Figure 6.** Normalized upper-TsVaRs as a function of the accuracy parameter $a$ at Tsurugi (red), Nakajima (green), and Kazarashi (blue).

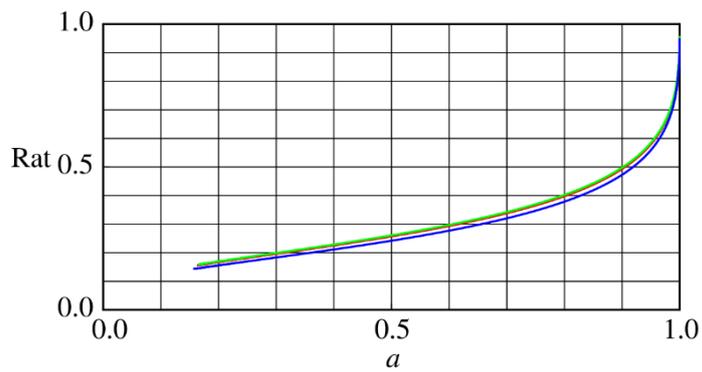

**Figure 7.** Normalized lower-TsVaRs as a function of the accuracy parameter $a$ at Tsurugi (red), Nakajima (green), and Kazarashi (blue).

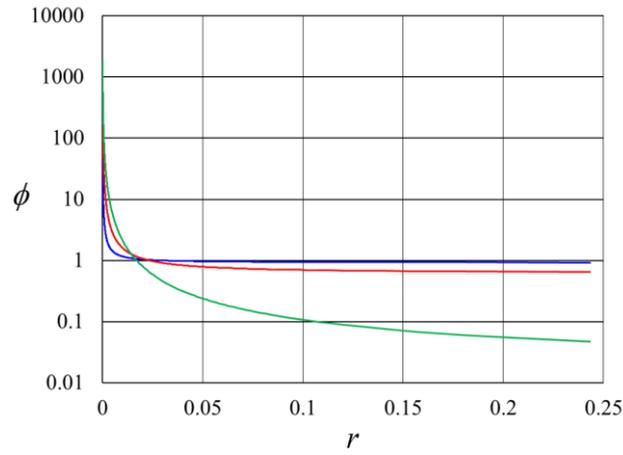

**Figure 8.** Worst-case Radon–Nikodym derivatives at Kazarashi for the overestimation case: $a = 0.99$ (blue), $a = 0.91$ (red), and $a = 0.50$ (green).

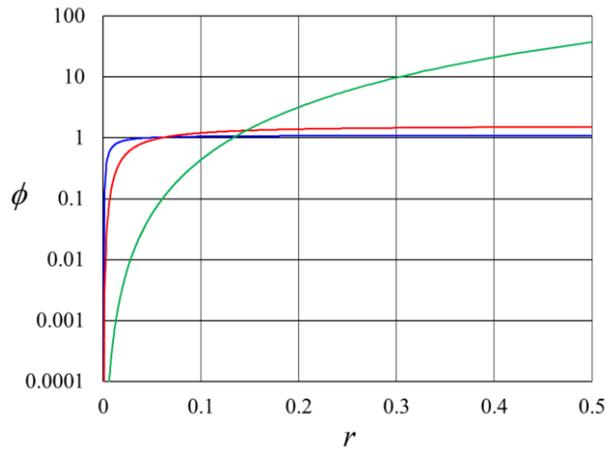

**Figure 9.** Worst-case Radon–Nikodym derivatives at Kazarashi for the underestimation case: $a = 0.99$ (blue), $a = 0.91$ (red), and $a = 0.19$ (green).

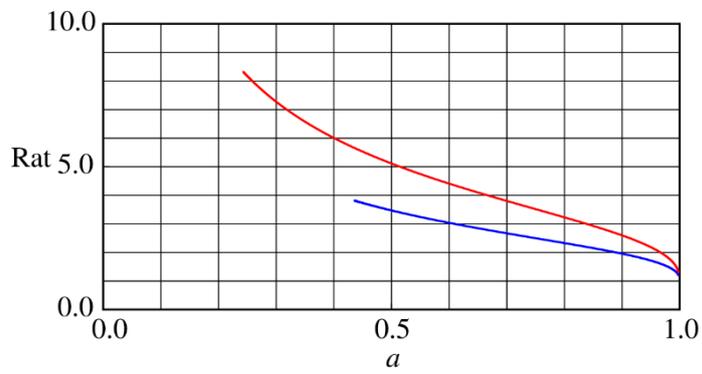

**Figure 10.** $q$-dependence of the normalized upper-TsVaR at Kazarashi: $q = 0.33$ (blue) and $q = 0.37$ (red).

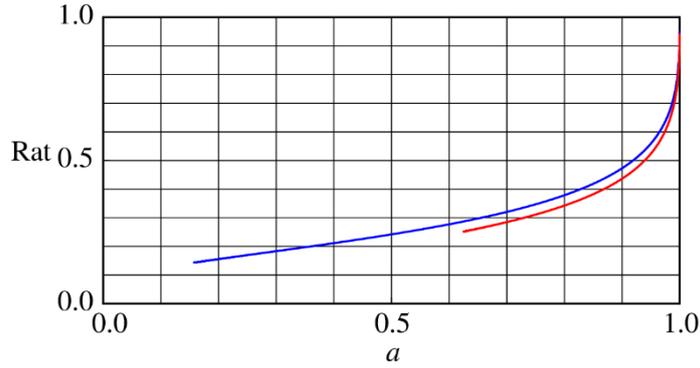

**Figure 11.** $q$-dependence of the normalized lower-TsVaR at Kazarashi: $q = 1.33$ (blue) and $q = 2.00$ (red).

**4. Conclusion**

We proposed a novel methodology based on TsVaRs for the statistical evaluation of long-memory processes subject to ambiguity of the reversion measure. We demonstrated that the TsVaRs serve as a more flexible risk measure that can cover a wider range of cases than the conventional EVaR. The semi-implicit gradient descent method is easily implementable, and the duality relation between the TsVaRs and corresponding nonlinear expectations enabled obtaining the distorted reversion measure under worst-case ambiguities. The application of the proposed methodology to real-world data demonstrated the effectiveness of the TsVaRs in evaluating the key statistics of the streamflow discharge.

We assumed that the reversion measure is the source of model ambiguity. However, the supOU process also has the Lévy measure, which may induce model ambiguity as well. TsVaRs may not be applicable to the ambiguity of the Lévy measure because they are based on the distortion of a probability measure. The Lévy measure is integrable only for compound Poisson processes that do not cover the tempered stable processes considered in our application. Applications of the TsVaR to other environmental processes such as the expansion/contraction of wetlands (Bertassello et al., 2022) and coastal flood risk assessment (Ha et al., 2022) would be interesting topics. The TsVaRs can also be applied to problems other than streamflow environments where long-memory processes are applicable, such as carbon dioxide emissions related to climate change (Claudio-Quiroga and Gil-Alana, 2022; Proietti and Maddanu, 2022) and the earth surface temperature (Beaulieu et al., 2019). A key future topic will be an extension of the proposed methodology to problems with an explicit modeling of the seasonality and diel variations found in streamflow environments (Jackson et al., 2022). The nonlinear expectations utilized in this paper are interesting by themselves. The authors are currently investigating their well-posedness as well as ill-posedness and their implications to the ambiguity aversion of environmental managers (Yoshioka and Yoshioka, 2023).

**References**

Ahmadi-Javid, A. (2012). Entropic value-at-risk: A new coherent risk measure. *Journal of Optimization Theory and Applications*, *155*(3), 1105–1123. doi:10.1007/s10957-011-9968-2

Anderson, E. W., Hansen, L. P., & Sargent, T. J. (2003). A quartet of semigroups for model specification, robustness, prices of risk, and model detection. *Journal of the European Economic Association*, *1*(1), 68–123. doi:10.1162/154247603322256774

Balter, A. G., & Pelsser, A. (2021). Quantifying ambiguity bounds via time-consistent sets of indistinguishable models. *Systems and Control Letters*, *149*, 104877. doi:10.1016/j.sysconle.2021.104877

Barndorff-Nielsen, O. E. (2001). Superposition of Ornstein–Uhlenbeck type processes. *Theory of Probability and its Applications*, *45*(2), 175–194. doi:10.1137/S0040585X97978166

Barndorff-Nielsen, O. E., & Stelzer, R. (2011). Multivariate supOU processes. *Annals of Applied Probability*, *21*(1), 140–182. doi:10.1214/10-AAP690

Beaulieu, C., Killick, R., Ireland, D., & Norwood, B. (2020). Considering long-memory when testing for changepoints in surface temperature: A classification approach based on the time-varying spectrum. *Environmetrics*, *31*(1), e2568. doi: 10.1002/env.2568

Beran, J., Feng, Y., Ghosh, S., & Kulik, R. (2013). *Long-memory processes*. Berlin, Heidelberg: Springer-Verlag.

Beran, J., Liu, H., & Ghosh, S. (2020). On aggregation of strongly dependent time series. *Scandinavian Journal of Statistics*, *47*(3), 690–710. doi:10.1111/sjos.12421

Bertassello, L. E., Durighetto, N., & Botter, G. (2022). Eco-hydrological modelling of channel network dynamics—part 2: application to metapopulation dynamics. *Royal Society Open Science*, *9*(11), 220945. doi:10.1098/rsos.220945

Bertassello, L. E., Jawitz, J. W., Bertuzzo, E., Botter, G., Rinaldo, A., Aubeneau, A. F., Hoverman, J. T., & Rao, P. S. C. (2022). Persistence of amphibian metapopulation occupancy in dynamic wetlandscapes. *Landscape Ecology*, *37*(3), 695–711. doi:10.1007/s10980-022-01400-

Cao, Y., Lu, J., & Lu, Y. (2019). Exponential decay of Rényi divergence under Fokker–Planck equations. *Journal of Statistical Physics*, *176*(5), 1172–1184. doi:10.1007/s10955-019-02339-8

Chen, S., & Wang, J. (2022). An integrative analytical framework and evaluation system of water environment security in the context of agricultural non-point source perspective. *Environmental Research Communications*. doi:10.1088/2515-7620/acabb5

Chen, Y., Wang, Z., & Zhang, Z. (2019). Mark to market value at risk. *Journal of Econometrics*, *208*(1), 299–321. doi:10.1016/j.jeconom.2018.09.017

Claudio-Quiroga, G., & Gil-Alana, L. A. (2022). CO2 emissions persistence: Evidence using fractional integration. *Energy Strategy Reviews*, *43*, 100924. doi:10.1016/j.esr.2022.100924

Dang, M. H., Umeda, S., & Yuhi, M. (2014). Long-term riverbed response of lower Tedori River, Japan, to sediment extraction and dam construction. *Environmental Earth Sciences*, *72*(8), 2971–2983. doi:10.1007/s12665-014-3202-0

Datry, T., Truchy A., Olden, J. D., Busch, M. H., Stubbington, R., Dodds, W. K., Zipper, S., Yu, S., Messager, M. L., Tonkin, J. D., Kaiser, K. E., Hammond, J. C., Moody, E. K., Burrows, R. M., Sarremejane, R., DelVecchia, A. G., Fork, M/ L/. Little, C. J., Walker, R. H., Walters, A. W., & Allen, D. (2022). Causes, responses, and implications of anthropogenic versus natural flow intermittence in river networks. *BioScience*. doi:10.1093/biosci/biac098

Diamond, J. S., Pinay, G., Bernal, S., Cohen, M. J., Lewis, D., Lupon, A., Zarnetske, J., & Moatar, F. (2022). Light and hydrologic connectivity drive dissolved oxygen synchrony in stream networks. *Limnology and Oceanography*. doi:10.1002/lno.12271

Dowd, K., Cotter, J., & Sorwar, G. (2008). Spectral risk measures: Properties and limitations. *Journal of Financial Services Research*, *34*(1), 61–75. doi:10.1007/s10693-008-0035-6

Durighetto, N., Bertassello, L. E., & Botter, G. (2022). Eco-hydrological modelling of channel network dynamics—part 1: stochastic simulation of active stream expansion and retraction. *Royal Society Open Science*, *9*(11), 220944. doi:10.1098/rsos.220944

Fasen, V., & Klüppelberg, C. (2007). Extremes of supOU processes. In *Stochastic Analysis and Applications*. Berlin, Heidelberg: Springer, (339–359). doi:10.1007/978-3-540-70847-6_14

Gupta, A., & Govindaraju, R. S. (2023). Uncertainty quantification in watershed hydrology: Which method to use? *Journal of Hydrology*, 616, 128749. doi:10.1016/j.jhydrol.2022.128749

Ha, S., Fujimi, T., Jiang, X., Mori, N., Begum, R. A., Watanabe, M., Tatano, H., & Nakakita, E. (2022). Estimating household preferences for coastal flood risk mitigation policies under ambiguity. *Earth's Future*, 10(12), EF003031, e2022. doi:10.1029/2022EF003031

Hainaut, D. (2022). *Continuous time processes for finance. Switching, self-exciting, fractional and other recent dynamics*. Cham, Germany: Springer.

Herrera, P. A., Marazuela, M. A., & Hofmann, T. (2022). Parameter estimation and uncertainty analysis in hydrological modeling. *WIREs Water*, 9(1), e1569. doi:10.1002/wat2.1569

Hough, I., Moggridge, H., Warren, P., & Shucksmith, J. (2022). Regional flow–ecology relationships in small, temperate rivers. *Water and Environment Journal*, 36(1), 142–160. doi:10.1111/wej.12757

Hrafnkelsson, B., Sigurdarson, H., Rögnvaldsson, S., Jansson, A. Ö., Vias, R. D., & Gardarsson, S. M. (2022). Generalization of the power-law rating curve using hydrodynamic theory and Bayesian hierarchical modeling. *Environmetrics*, 33(2), e2711. doi:10.1002/env.2711

Imai, J. (2022). A numerical method for hedging Bermudan options under model uncertainty. *Methodology and Computing in Applied Probability*, 24(2), 893–916. doi:10.1007/s11009-021-09901-6

Iwasaki Yoshioka, Y., Nakamura, K., Nakano, T., Horino, H., Shin, K. C., Hashimoto, S., & Kawashima, S. (2016). Multiple-indicator study of groundwater flow and chemistry and the impacts of river and paddy water on groundwater in the alluvial fan of the Tedori River, Japan. *Hydrological Processes*, 30(16), 2804–2816. doi:10.1002/hyp.10785

Jackson, S., Anderson, E. P., Piland, N. C., Carriere, S., Java, L., & Jardine, T. D. (2022). River rhythmicity: A conceptual means of understanding and leveraging the relational values of rivers. *People and Nature*, 4(4), 949–962. doi:10.1002/pan3.10335

Kimura, M., Kobayashi, S., Mitsuyasu, M., Xie, W., & Iida, T. (2022). Simulation model of water temperature variation in dual-purpose canals considering return flow from upstream paddy fields. *Irrigation and Drainage*, 71(S1)(Suppl. 1), 138–154. doi:10.1002/ird.2697

Kumbhakar, M., & Tsai, C. W. (2022). A probabilistic model on streamwise velocity profile in open channels using Tsallis relative entropy theory. *Chaos, Solitons and Fractals*, 165, 112825. doi:10.1016/j.chaos.2022.112825

Laroche, C., Olteanu, M., & Rossi, F. (2022). Pesticide concentration monitoring: Investigating spatio-temporal patterns in left censored data. *Environmetrics*, e2756. doi:10.1002/env.2756

Leleux, P., Courtain, S., Guex, G., & Saerens, M. (2021). Sparse randomized shortest paths routing with Tsallis divergence regularization. *Data Mining and Knowledge Discovery*, 35(3), 986–1031. doi:10.1007/s10618-021-00742-y

Ma, H., & Tian, D. (2021). Generalized entropic risk measures and related BSDEs. *Statistics and Probability Letters*, 174, 109110. doi:10.1016/j.spl.2021.109110

Mastrantonio, G., Jona Lasinio, G., Pollice, A., Teodonio, L., & Capotorti, G. (2022). A Dirichlet process model for change-point detection with multivariate bioclimatic data. *Environmetrics*, 33(1), e2699. doi:10.1002/env.2699

Mudelsee, M. (2007). Long memory of rivers from spatial aggregation. *Water Resources Research*, 43(1). doi:10.1029/2006WR005721

Nallaperuma, B., & Asaeda, T. (2020). The long-term legacy of riparian vegetation in a hydrogeomorphologically remodelled fluvial setting. *River Research and Applications*, 36(8), 1690–1700. doi:10.1002/rra.3665

Pedersen, J. (2003). *The Lévy-Ito decomposition of an independently scattered random measure. MaPhySto, department of mathematical sciences*. Aarhus: University of Aarhus.

Phillips, J. (2022). A rapid sustainability dynamic assessment of the USA and China 1995–2018. *Environmental Monitoring and Assessment*, 194(7), 490. doi:10.1007/s10661-022-10141-5

Proietti, T., & Maddanu, F. (2022). Modelling cycles in climate series: The fractional sinusoidal waveform process. Modeling Cycles In Climate Series. *Journal of Econometrics*. doi:10.1016/j.jeconom.2022.04.008

Qiu, R., Wang, D., Singh, V. P., Zhang, H., Tao, Y., Wu, J., & Wang, Y. (2022). Ecological responses of spawning habitat suitability to changes in flow and thermal regimes influenced by hydropower operation. *Ecohydrology*, e2507. doi:10.1002/eco.2507

Randrianambinina, S. A., & Esunge, J. (2022). Applications of a superposed Ornstein-Uhlenbeck type processes. *Journal of Stochastic Analysis*, 3(4), 3. doi:10.31390/josa.3.4.03

Ranjram, M., & Craig, J. R. (2022). Upscaling hillslope-scale subsurface flow to inform catchment-scale recession behavior. *Water Resources Research*, 58(10), WR031913, e2021. doi:10.1029/2021WR031913

Rinaldo, A., & Rodriguez-Iturbe, I. (2022). Ecohydrology 2.0. *Rendiconti Lincei. Scienze Fisiche e Naturali*, 33(2), 245–270. doi:10.1007/s12210-022-01071-y

Steffy, L. Y., & Shank, M. K. (2018). Considerations for using turbidity as a surrogate for suspended sediment in small, ungaged streams: Time-series selection, streamflow estimation, and regional transferability. *River Research and Applications*, 34(10), 1304–1314. doi:10.1002/rra.3373

Takase, K., & Fujihara, Y. (2019). Evaluation of the effects of irrigation water on groundwater budget by a hydrologic model. *Paddy and Water Environment*, 17(3), 439–446. doi:10.1007/s10333-019-00739-w

Tsang, M. Y., Sit, T., & Wong, H. Y. (2022). Robust portfolio optimization with respect to spectral risk measures under correlation uncertainty. *Applied Mathematics and Optimization*, 191, 47–77. doi:10.1007/s10107-018-1347-4

Vera-Valdés, J. E. (2020). On long memory origins and forecast horizons. *Journal of Forecasting*, 39(5), 811–826. doi:10.1002/for.2651

Wang, J., & Guo, Y. (2019). Stochastic analysis of storm water quality control detention ponds. *Journal of Hydrology*, 571, 573–584. doi:10.1016/j.jhydrol.2019.02.001

Wasserburger, A., Hametner, C., & Didcock, N. (2020). Risk-averse real driving emissions optimization considering stochastic influences. *Engineering Optimization*, 52(1), 122–138. doi:10.1080/0305215X.2019.1569646

Watanabe, R., Ibuki, T., Sakayanagi, Y., Funada, R., & Sampei, M. (2022). Risk-aware energy management for drive mode control in plug-in hybrid electric vehicles. *IEEE Access*, 10, 103619–103631. doi:10.1109/ACCESS.2022.3206091

Yoshioka, H. (2021). Fitting a superposition of Ornstein–Uhlenbeck process to time series of discharge in a perennial river environment. *ANZIAM Journal*, 63, C84–C96. doi:10.21914/anziamj.v63.16985

Yoshioka, H., Tanaka, T., Yoshioka, Y., & Hashiguchi, A. (2023). Stochastic optimization of a mixed moving average process for controlling non-Markovian streamflow environments. *Applied Mathematical Modelling*, 116, 490–509. doi:10.1016/j.apm.2022.11.009

Yoshioka, H., & Yoshioka, Y. (2022). Stochastic streamflow and dissolved silica dynamics with application to the worst-case long-run evaluation of water environment. *Optimization and Engineering*. doi:10.1007/s11081-022-09743-2

Yoshioka, H., & Yoshioka, Y. (2023).The expectation of a mixed moving average process subject to ambiguous Lévy basis, ASMDA 2023 June 6-9, 2023, Crete, Greece. (To appear).

Yu, H., & Sun, J. (2021). Robust stochastic optimization with convex risk measures: A discretized subgradient scheme. *Journal of Industrial and Management Optimization*, 17(1), 81–99. doi:10.3934/jimo.2019100

Zeng, Y., Liu, D., Guo, S., Xiong, L., Liu, P., Yin, J., & Wu, Z. (2022). A system dynamic model to quantify the impacts of water resources allocation on water–energy–food–society (WEFS) nexus. *Hydrology and Earth System Sciences*, 26(15), 3965–3988. doi:10.5194/hess-26-3965-2022

Zou, Z., Xia, Z., & Hu, T. (2022). Tsallis value-at-risk: Generalized entropic value-at-risk. *Probability in the Engineering and Informational Sciences*, 1–20. doi:10.1017/S0269964822000444

**Supporting information of "Statistical evaluation of a long-memory process using the generalized entropic Value-at-Risk".**

**Appendix A. Mathematical results on the main text**

**A.1 Proofs of Propositions in the main text**

*Proof of Proposition 1*

For $q \neq 1$, the condition (18) is satisfied when

$$\int_0^{+\infty} \left(1 + (1-q)\frac{\lambda_0}{r}\right)^{\frac{1}{1-q}} \pi(dr) < +\infty \quad \text{with some} \quad \lambda_0 > 0. \tag{A44}$$

Firstly, as the integrand must be defined for all $r > 0$, we need $0 < q \leq 1$. For $q < 1$, then we must have the integrability of $r^{\frac{-1}{1-q}} \pi(dr)$ for small $r > 0$, and hence $0 < q < 1 - \alpha^{-1}$. In this case, the inequality in (18) is satisfied for all $\lambda_0 > 0$. This integrability is not satisfied for $1 - \alpha^{-1} \leq q < 1$ because the inequality in (18) fails for all $\lambda_0 > 0$. Finally, the case $q = 1$ is excluded because $\exp\left(\frac{\lambda_0}{r}\right) \pi(dr)$ is not integrable near $r = 0$ for all $\lambda_0 > 0$. In summary, it follows that the condition (18) is satisfied if and only if $0 < q < 1 - \alpha^{-1}$, completing the proof.

□

*Proof of Proposition 2*

The convexity of the upper-TsVaR and $\bar{g}$ is the direct consequence of Lemma 3.2, Lemma 3.3, and Proposition 3.4 of Zou et al. (2022). The statement of the second sentence can be proven by a contradiction argument. On the one hand, we have $0 < \bar{g}(\lambda_0) < +\infty$ for some $\lambda_0 > 0$. On the other hand, we have

$$\begin{aligned}\bar{g}(\lambda^{-1}) &= \lambda \ln_q \left(\frac{1}{a}\int_0^{+\infty} \exp_q\left(\frac{1}{\lambda r}\right)\pi(dr)\right) \\ &\geq \lambda \ln_q \left(\frac{1}{a}\int_0^{+\infty} \pi(dr)\right) \quad \text{for all} \quad \lambda > 0. \tag{A45}\\ &\geq \lambda \ln_q \left(\frac{1}{a}\right)\end{aligned}$$

We then have

$$\liminf_{\lambda \to +\infty} \bar{g}(\lambda^{-1}) \geq \liminf_{\lambda \to +\infty} \lambda \ln_q \left(\frac{1}{a}\right) = +\infty. \tag{A46}$$

Therefore, the infimum of (17) is attained in some compact set because of the convexity stated in the first sentence of the proposition, completing the proof.

□

*Proof of Proposition 3*

For $q \neq 1$, the condition (18) is satisfied if

$$\int_0^{+\infty} \left(1+(q-1)\frac{\lambda_0}{r}\right)^{\frac{-1}{q-1}} \pi(dr) = \int_0^{+\infty} \frac{r^{\frac{1}{q-1}}}{(r+(q-1)\lambda_0)^{\frac{1}{q-1}}} \pi(dr) < +\infty \quad \text{with some } \lambda_0 > 0. \quad (A47)$$

Firstly, as the integrand must be defined for all $r > 0$, we must have $q \geq 1$. Then, for each $\lambda_0 > 0$, it suffices to see

$$\int_0^{+\infty} \frac{r^{\frac{1}{q-1}}}{(r+(q-1)\lambda_0)^{\frac{1}{q-1}}} \pi(dr) \leq \int_0^{+\infty} \pi(dr) = 1. \quad (A48)$$

□

*Proof of Proposition 5*

The left and middle inequalities of (24) are immediate, while the right inequality of (24) follows from the Jensen's inequality

$$\int_0^{+\infty} \exp_q\left(\frac{\lambda}{r}\right) \pi(dr) \geq \exp_q\left(\int_0^{+\infty} \frac{\lambda}{r} \pi(dr)\right) \leftrightarrow \frac{1}{\lambda} \ln_q\left(\int_0^{+\infty} \exp_q\left(\frac{\lambda}{r}\right) \pi(dr)\right) \geq \int_0^{+\infty} \frac{1}{r} \pi(dr) = R. \quad (A49)$$

and the arbitrariness of $\lambda > 0$. The inequalities in (25) are proven in the same way by noticing that

$$-\ln_q\left(\frac{1}{a}\int_0^{+\infty} \exp_q\left(-\frac{\lambda}{r}\right) \pi(dr)\right) \leq -\ln_q\left(\frac{1}{b}\int_0^{+\infty} \exp_q\left(-\frac{\lambda}{r}\right) \pi(dr)\right) \leq -\ln_q\left(\int_0^{+\infty} \exp_q\left(-\frac{\lambda}{r}\right) \pi(dr)\right) \leq \lambda R$$

(A50)

for any $\lambda > 0$.

□

*Proof of Proposition 6*

The proof of the first equality (31) follows from the direct method of calculus of variations that the right-hand side of (31) is minimized by an analytical calculation. It is straightforward to check that, as in Hansen and Miao (2018), $\lambda \int_0^{+\infty} \{\phi(r)\}^q r^{-1} \pi(dr) + H_q(\phi)$ is strictly convex with respect to $\phi \in \mathfrak{F}$ and hence we obtain that the (unique) minimizer $\phi_*$ has the form

$$\phi_*(r) = C_* \exp_q\left(-\lambda r^{\frac{-1}{1-q}}\right), \quad (A51)$$

and the constant $C_* > 0$ is uniquely determined as

$$C_* = \left(\int_0^{+\infty} \exp_q\left(-\lambda r^{\frac{-1}{1-q}}\right) \pi(dr)\right)^{-1}. \quad (A52)$$

Substituting (A51)-(A52) into the right-hand side of (31) yields its left-hand, and the first equality is proven. Note that the integral (A52) exists by the assumption $q \geq 1$.

The proof of the second equality (32) follows the lines of the Proof of Theorem 3.8 of Zou et al. (2022) with an adaptation to our setting by considering that the lower-TsVaR is given by a maximization problem. We show the proof for $q > 1$, and that for $q = 1$ follows by formally taking the limit $q \to 1$ in the argument below. By the definition of the lower-TsVaR and the $\ln_q$ function, we have

$$\begin{aligned}\underline{\text{TsVaR}}_a &= \sup_{\lambda>0}\left\{-\frac{1}{\lambda}\ln_q\left(\frac{1}{a}\int_0^{+\infty}\exp_q\left(-\frac{\lambda}{r}\right)\pi(\mathrm{d}r)\right)\right\}\\ &= \sup_{\lambda>0}\left\{-\frac{1}{\lambda}\left[a^{q-1}\ln_q\left(\int_0^{+\infty}\exp_q\left(-\frac{\lambda}{r}\right)\pi(\mathrm{d}r)\right)+\ln_q\left(\frac{1}{a}\right)\right]\right\}\\ &= \sup_{\lambda>0}\frac{1}{\lambda}\left(-a^{q-1}\ln_q\left(\int_0^{+\infty}\exp_q\left(-\frac{\lambda}{r}\right)\pi(\mathrm{d}r)\right)-\ln_q\left(\frac{1}{a}\right)\right)\\ &= \sup_{\lambda>0}\frac{1}{\lambda}\left(-a^{q-1}\ln_q\left(\int_0^{+\infty}\exp_q\left(-\frac{\lambda}{r}\right)\pi(\mathrm{d}r)\right)+a^{q-1}\ln_q(a)\right)\end{aligned} \quad (A53)$$

Substituting (32) into (A53) obtains, with the help of the classical technique of the Lagrangian multiplier, that

$$\begin{aligned}\underline{\text{TsVaR}}_a &= \sup_{\lambda>0}\frac{1}{\lambda}\left(-a^{q-1}\left[-\ln_q\left(\int_0^{+\infty}\exp_q\left(-\frac{\lambda}{r}\right)\pi(\mathrm{d}r)\right)\right]+a^{q-1}\ln_q(a)\right)\\ &= \sup_{\lambda>0}\frac{1}{\lambda}\left(a^{q-1}\inf_{\phi\in\mathfrak{F}'}\left\{\lambda\int_0^{+\infty}\{\phi(r)\}^q\frac{1}{r}\pi(\mathrm{d}r)+H_q(\phi)\right\}+a^{q-1}\ln_q(a)\right)\\ &= \sup_{\lambda>0}\left(a^{q-1}\inf_{\phi\in\mathfrak{F}'}\left\{\int_0^{+\infty}\{\phi(r)\}^q\frac{1}{r}\pi(\mathrm{d}r)+\frac{1}{\lambda}H_q(\phi)\right\}+\frac{1}{\lambda}a^{q-1}\ln_q(a)\right),\\ &= \sup_{\lambda>0}\inf_{\phi\in\mathfrak{F}'}\left\{a^{q-1}\int_0^{+\infty}\{\phi(r)\}^q\frac{1}{r}\pi(\mathrm{d}r)+\frac{a^{q-1}}{\lambda}\left(H_q(\phi)+\ln_q(a)\right)\right\}\\ &= \inf_{\phi\in\mathfrak{F}}\left\{a^{q-1}\int_0^{+\infty}\{\phi(r)\}^q\frac{1}{r}\pi(\mathrm{d}r)\right\}\end{aligned} \quad (A54)$$

which is the desired result.

□

### A.2 The strict convexity result

The following strict convexity result mentioned at the end of **Remark 4** is proven here.

**Proposition A.1** Assume $0 < q < 1 - \alpha^{-1}$. Then, the function $g(\lambda) \equiv \bar{g}(\lambda^{-1})$ is strictly convex for $\lambda > 0$.

*Proof of Proposition 4.1*

Set

$$I_k \equiv \int_0^{+\infty}\left(\frac{1}{r}\right)^{k-1}\left\{\exp_q\left(\frac{1}{\lambda r}\right)\right\}^{d(k)}\pi(\mathrm{d}r) < +\infty \quad (A55)$$

with $d(1) = 1$, $d(2) = q$, $d(3) = 2q - 1$. A straightforward calculation shows

$$\frac{dI_1}{d\lambda} = -\frac{1}{\lambda^2} I_2 \quad \text{and} \quad \frac{dI_2}{d\lambda} = -\frac{q}{\lambda^2} I_3, \tag{A56}$$

which will be used in the sequel. Each $I_k$ is well-defined because their integrands commonly behave for small $r > 0$ as the integrable function $r^{-1/(1-q)}$ with respect to $\pi$.

It is sufficient to show $\frac{d^2}{d\lambda^2} g(\lambda) > 0$ for all $\lambda > 0$. By (39), we have the first-order derivative

$$\frac{d}{d\lambda} g(\lambda) = -\ln_q\left(\frac{1}{a} I_1\right) + \frac{1}{a\lambda} I^2 \left(\frac{1}{a} I_1\right)^{-q}. \tag{A57}$$

We also have

$$\frac{d}{d\lambda}\left\{-\ln_q\left(\frac{1}{a} I_1\right)\right\} = -\left(\frac{1}{a} I_1\right)^{-q} \frac{d}{d\lambda}\left\{\frac{1}{a} I_1\right\}$$

$$= -\left(\frac{1}{a} I_1\right)^{-q} \frac{1}{a}\left(-\frac{1}{\lambda^2} I_2\right) \tag{A58}$$

$$= \frac{1}{\lambda^2}\left(\frac{1}{a}\right)^{1-q} I_2 (I_1)^{-q}$$

and

$$\frac{d}{d\lambda}\left\{\frac{1}{a\lambda} I_2 \left(\frac{1}{a} I_1\right)^{-q}\right\}$$

$$= \left(\frac{1}{a}\right)^{1-q}\left[-\frac{1}{a\lambda^2} I_2 (I_1)^{-q} + \frac{1}{a\lambda}\frac{d}{d\lambda}\left\{I_2 (I_1)^{-q}\right\}\right]$$

$$= \left(\frac{1}{a}\right)^{1-q}\left[-\frac{1}{a\lambda^2} I_2 (I_1)^{-q} + \frac{1}{a\lambda}\frac{dI_2}{d\lambda}(I_1)^{-q} - \frac{1}{a\lambda^3} q (I_1)^{-q-1}\frac{dI_1}{d\lambda} I_2\right] \tag{A59}$$

$$= \left(\frac{1}{a}\right)^{1-q}\left[-\frac{1}{a\lambda^2} I_2 (I_1)^{-q} - \frac{q}{a\lambda^3} I_3 (I_1)^{-q} + \frac{1}{a\lambda^3} q (I_1)^{-q-1} (I_2)^2\right]$$

We then obtain

$$\frac{d^2}{d\lambda^2} g(\lambda) = -\left(\frac{1}{a}\right)^{1-q}\left[\frac{1}{\lambda^2} I_2 (I_1)^{-q} - \frac{1}{\lambda^2} I_2 (I_1)^{-q} - \frac{q}{\lambda^3} I_3 (I_1)^{-q} + \frac{1}{\lambda^3} q (I_1)^{-q-1} (I_2)^2\right], \tag{A60}$$

$$= \frac{q}{\lambda^3}\left(\frac{1}{a}\right)^{1-q}\left(I_1 I_3 - (I_2)^2\right)$$

where

$$I_1 I_3 - (I_2)^2$$

$$= \left(\int_0^{+\infty} \frac{1}{r^2}\left\{\exp_q\left(\frac{1}{\lambda r}\right)\right\}^{2q-1} \pi(dr)\right)\left(\int_0^{+\infty} \exp_q\left(\frac{1}{\lambda r}\right) \pi(dr)\right) - \left(\int_0^{+\infty} \frac{1}{r}\left\{\exp_q\left(\frac{1}{\lambda r}\right)\right\}^q \pi(dr)\right)^2. \tag{A61}$$

Due to

$$\int_0^{+\infty} \frac{1}{r^2} \left\{ \exp_q\left(\frac{1}{\lambda r}\right) \right\}^{2q-1} \times \exp_q\left(\frac{1}{\lambda r}\right) = \left[ \frac{1}{r} \left\{ \exp_q\left(\frac{1}{\lambda r}\right) \right\}^q \right]^2, \tag{A62}$$

we can use the multivariate Jensen inequality for the concave bi-variate (Proposition C.1 of Marshall et al. (2010)) to the concave function $f(x,y) = \sqrt{xy}$ ($x, y \geq 0$) with the fact that $f$ in not affine; we have $I_1 I_3 - (I_2)^2 > 0$ for all $\lambda > 0$, and hence the strict convexity of $g$ from (A60).

□

*Remark A.1* Similar strict convexity holds true for the lower-TsVaR case.

## References

Hansen, L. P., & Miao, J. (2018). Aversion to ambiguity and model misspecification in dynamic stochastic environments. Proceedings of the National Academy of Sciences of the United States of America, 115(37), 9163–9168. doi:10.1073/pnas.1811243115

Marshall, A. W., Olkin, I., & Arnold, B. C. (2010). Inequalities: Theory of majorization and its applications. *Springer Series in Statistics*. New York: Springer.

Zou, Z., Xia, Z., & Hu, T. (2022). Tsallis value-at-risk: Generalized entropic value-at-risk. *Probability in the Engineering and Informational Sciences*, 1–20. doi:10.1017/S0269964822000444